\documentclass[useAMS,usenatbib]{mn2e}
\usepackage{graphicx,psfig,cite,epsf}  
\usepackage{times}
\usepackage{subfigure}
\usepackage{textcomp}

\title{X-ray spectral states and metallicity in the ultraluminous X-ray sources NGC 1313 X-1 and X-2}
\author[Fabio Pintore \& Luca Zampieri]{Fabio Pintore$^{1,2}$ \& Luca Zampieri$^{2}$\\
$^1$ Dipartimento di Astronomia, Universit\`a di Padova, Vicolo dell'Osservatorio 3, I-35122 Padova, Italy\\
$^2$ INAF-Osservatorio Astronomico di Padova, Vicolo dell'Osservatorio 5, I-35122 Padova, Italy} 

\begin{document}

\date{Accepted ... Received ...; in original form ...}
\pagerange{\pageref{firstpage}--\pageref{lastpage}} \pubyear{2010}
\maketitle

\label{firstpage}

\begin{abstract}
We present a systematic analysis of the X-ray spectra of NGC 1313 X-1 and NGC 1313 X-2, 
using three years of {\it XMM-Newton} observations. We fitted
the continuum with a Comptonization model plus a multicolor blackbody disc, 
that describes the effects of an accretion disc plus a corona. We checked the consistency of this spectral model on the
basis of the variability patterns of its spectral parameters. We found
that the two sources show different spectral states. We tentatively interpret 
the observed behaviour of NGC 1313 X-1 and X-2 within the framework of 
near Eddington and/or super-Eddington accretion. We also attempted to determine the
chemical abundances in the local environment of NGC 1313 X-1 and X-2 from the
EPIC and RGS spectra. The results appear to indicate subsolar metallicity for both sources. 
\end{abstract}


\section{Introduction}
\label{sect1}

Ultra luminous X-ray sources (ULXs) are extragalactic, point-like, off nuclear X-ray sources with observed isotropic
X-ray luminosities higher than the Eddington luminosity for accretion onto a 10 $M_\odot$ black hole, 
and typically in the range from $10^{39}$ erg s$^{-1}$ up to $\sim 10^{41}$ erg s$^{-1}$ (e.g. \citealt{fabbiano89}). 
Several pieces of observational evidence strongly suggest that the majority of ULXs 
are accreting black hole (BH) X-ray binaries (XRBs) with massive donors (e.g. \citealt{zampieri09} and references therein).

At present, the size and the origin of the BHs hosted in ULXs is still a matter of debate. 
It has been proposed that ULXs are powered by a stellar mass BHs (10$\div$20 $M_{\odot}$) accreting largely
above the Eddington limit and/or having beamed emission (e.g. \citealt{king01,king09,begelman06,socrates06,poutanen07}). 
Alternatively, the compact object could simply be bigger, and the accretion 
would be in the usual sub-Eddington regime. In this case the compact object would be an intermediate mass black 
hole (IMBH) with a mass in excess of $100 M_\odot$ (e.g. \citealt{colbert99}). Another interpretation follows from the possibility of forming massive stellar remnant BHs ($20\div 80$ $M_{\odot}$) from the direct collapse of low-metallicity, massive ($\ga 30-40 \, M_\odot$) stars (\citealt{fryer99,belczynski10}). Some ULXs may contain such massive stellar remnant BHs accreting at or slightly above the Eddington rate, formed in low metallicity environments ($Z\sim 0.1 Z_{\odot}$; e.g.\citealt{zampieri09,mapelli09,mapelli10}).

\begin{table*}
  \begin{center}
\footnotesize
   \caption{Log of the observations.}
   \label{ccd6}
   \begin{tabular}{l c c c c c c c}
\hline 
No. & Obs ID & Date & Exp$^a$ & Instr. (X-1)$^b$ & Instr. (X-2)$^b$ & Net counts(X-1) & Net counts(X-2)\\
& & & (ks) & & &  & \\
\hline
1 & 0150280101 & 11/25/2003 & 1 & M1/M2 & M1/M2/pn & 651, 637 & 578, 643, 734 \\
2 & 0150280301 & 12/21/2003 & 7.4 & pn & M1/M2/pn & 7648 & 2960, 2874, 6304 \\
3 & 0150280401 & 12/23/2003 & 3.2 & pn & M1/M2/pn & 2223 & 2215, 2151, 3030 \\
4 & 0150280501 & 12/25/2003 & 1.7 & pn & M1/M2/pn & 484 & 1379, 1378, 904 \\
5 & 0150280601 & 01/08/2004 & 6.5 & pn & M1/M2/pn & 5524 &  2714, 1694, 1690\\
6 & 0150281101 & 01/16/2004 & 2.7 & M1/M2/pn & M1/M2/pn & 1770, 1993, 2366  & 887, 713, 1013 \\
7 & 0205230201 & 05/01/2004 & 7.8$^c$ & M1/M2 & M1/M2 & 1310, 1489 & 700, 783 \\
8 & 0205230301 & 06/05/2004 & 8.7 & M1/M2/pn & M1/M2/pn & 5446, 5151, 8685 & 3800, 3798, 819 \\
9 & 0205230401 & 08/23/2004 & 3.8 & M1/M2/pn & M1/M2/pn & 2397, 2643, 2323 & 1252, 1316, 1173 \\
10 & 0205230501 & 11/23/2004 & 12.5 & M1/M2 & M1/M2/pn & 3013, 3179 & 1523, 1594, 4164 \\
11 & 0205230601 & 02/07/2005 & 9.0 & M1/M2/pn & M1/M2/pn & 5025, 2435, 2187 & 3702,3697, 8224 \\
12 & 0301860101 & 03/06/2006 & 17.2 & pn & M1/M2/pn & 3980 & 4043, 4374, 11426 \\
13 & 0405090101 & 10/16/2006 & 78.6 & M1/M2/pn & pn & 24277, 25141, 58116 & 53271	 \\
\hline
\hline
\end{tabular}
\end{center}
$^a$ GTI of EPIC-pn \\
$^b$ pn = EPIC-pn camera; M1/M2 = EPIC-MOS1/MOS2 camera \\
$^c$ GTI of EPIC-MOS
\end{table*}

The spectra of many ULXs are qualitatively similar to those of Galactic XRBs, and can be well reproduced by a
multicolour disc blackbody with a low characteristic temperature ($\sim 0.1-0.4$ keV), plus a power-law
continuum with a spectral index $\sim 1.5-2.5$ (e.g. \citealt{feng05}).  The high luminosity and low temperature
of the disc component were interpreted as supporting evidence for the existence of IMBHs
\citep{miller03,miller04}. However, in the last few years, {\it Chandra} and {\it XMM-Newton} observations
revealed new behaviours, showing the existence of X-ray spectra with rather peculiar properties. For some of the
brightest ULXs, equally acceptable fits of their spectra may be obtained with (physically) different models,
that suggest the presence of an optically thick corona, a fast ionized outflow or a slim disc (e.g.
\citealt{stobbart06,goncalves06,mizuno07}). Recently, \citet{gladstone09} have shown that disc plus Comptonized
corona models fit the data of the highest quality {\it XMM-Newton} spectra of several ULXs well, suggesting
that the corona is cool and optically thick.

In order to increase our understanding of ULXs and shed light on the mechanism at the origin of their powerful
emission, it is crucial to investigate the evolution of their accretion flow through the variability of their
X-ray spectra.  In this work we present a systematic analysis of the X-ray spectra of two ULXs in the galaxy NGC
1313, using three years of {\it XMM-Newton} observations. The barred spiral galaxy NGC 1313 hosts
three ULXs, one of which (NGC 1313 X-3) is a known supernova (SN 1978K) interacting with the circumstellar
medium. The other two ULXs, NGC 1313 X-1 and X-2 (X-1 and X-2 hereafter), are located in different positions,
close to the nucleus ($\sim 50"$) X-1 and at the outskirts of the host galaxy ($\sim 6'$) X-2. NGC 1313 has been
observed several times over the years by {\it XMM-Newton} and a sufficient number of X-ray spectra are now
available to attempt a characterization of the spectral variability of the sources hosted in it.

\citet{feng06} fitted a sequence of 12 {\it XMM-Newton} observations of X-1 and X-2 
with a power-law plus multicolor disc blackbody model and found an anti-correlation between the luminosity and the inner
temperature of the MCD component. For this reason they concluded that such component 
does not originate in a standard accretion disc.
The optical and X-ray variability of X-2 was also investigated by Mucciarelli et al. (2007) to 
constrain the properties of the donor star and the binary system. They found that
the power-law component hardens as the flux increases, opposite to what usually shown by 
Galactic BH XRBs.

Another crucial issue related to ULX formation are the properties of the environment in which ULXs
are embedded. Claims regarding the correlation of ULXs with low metallicity environments have been recently reported.
\citet{swartz08} found that within the Local Volume the 
specific ULX frequency decreases with host galaxy mass above $\sim 10^{8.5} M_{\odot}$, meaning that
smaller, lower metallicity systems have more ULXs per unit mass than larger galaxies. 
\citet{mapelli09} and \citet{zampieri09} suggested that at least a fraction of ULXs may be powered 
by massive stellar BHs formed from the direct collapse of low-metallicity massive stars.
Using binary synthesis calculations, \citet{linden10} proposed another interpretation, in which the 
number, the lifetime and (to a less extent) the luminosity of high mass XRBs are enhanced at low 
metallicities. Only few measurements of the metallicity in the ULX environment are available,
and results are not conclusive (see e.g. \citealt{zampieri09} and references therein). 
While optical observations provide probably the best means to perform such measurements, also 
the abundances inferred from the detection of K-shell photoionization 
edges of intermediate mass or heavy elements in the X-ray absorption spectrum of ULXs can be
a viable tool, if high signal-to-noise spectra are available. This was already attempted 
by \citet{winter07} for a sample of 14 ULXs with {\it XMM-Newton} spectra, obtaining values that match the solar 
abundance, but no further investigation, especially using the high resolution RGS spectra, has
been attempted since then.

In this work we try to characterize the spectral variability of X-1 and X-2 using all the available
{\it XMM-Newton} data. Recently, detailed spectral variability analyses were performed by \citet{feng09} 
 and \citet{kajava09} on {\it XMM-Newton} and {\it Chandra} data of some ULXs, by \citet{dewangan10} 
on the last two {\it XMM-Newton} observations of 
NGC 1313 X-1, and by \citet{vierdayanti10} on {\it XMM-Newton} and Swift data of Ho IX X-1. We will try to constrain also 
the metallicity of the absorbing gas towards X-1 and X-2 using the RGS and EPIC spectra of the
longest observation, and stacking together all the RGS observations. 
Some preliminary results of this investigation were reported in \citet{pintore11}. Because of some 
improvements in the present analysis, the results and interpretation reported here supersede 
our previous ones.
The plan of the paper is the following. In \S~\ref{sect2} we summarize the data selection and reduction
procedures, while in \S~\ref{sect3} and \S~\ref{sect4} we present our results and
discuss them in \S~\ref{sect5}.

\begin{figure*}
 \subfigure{\includegraphics[height=6.5cm,width=8.3cm]{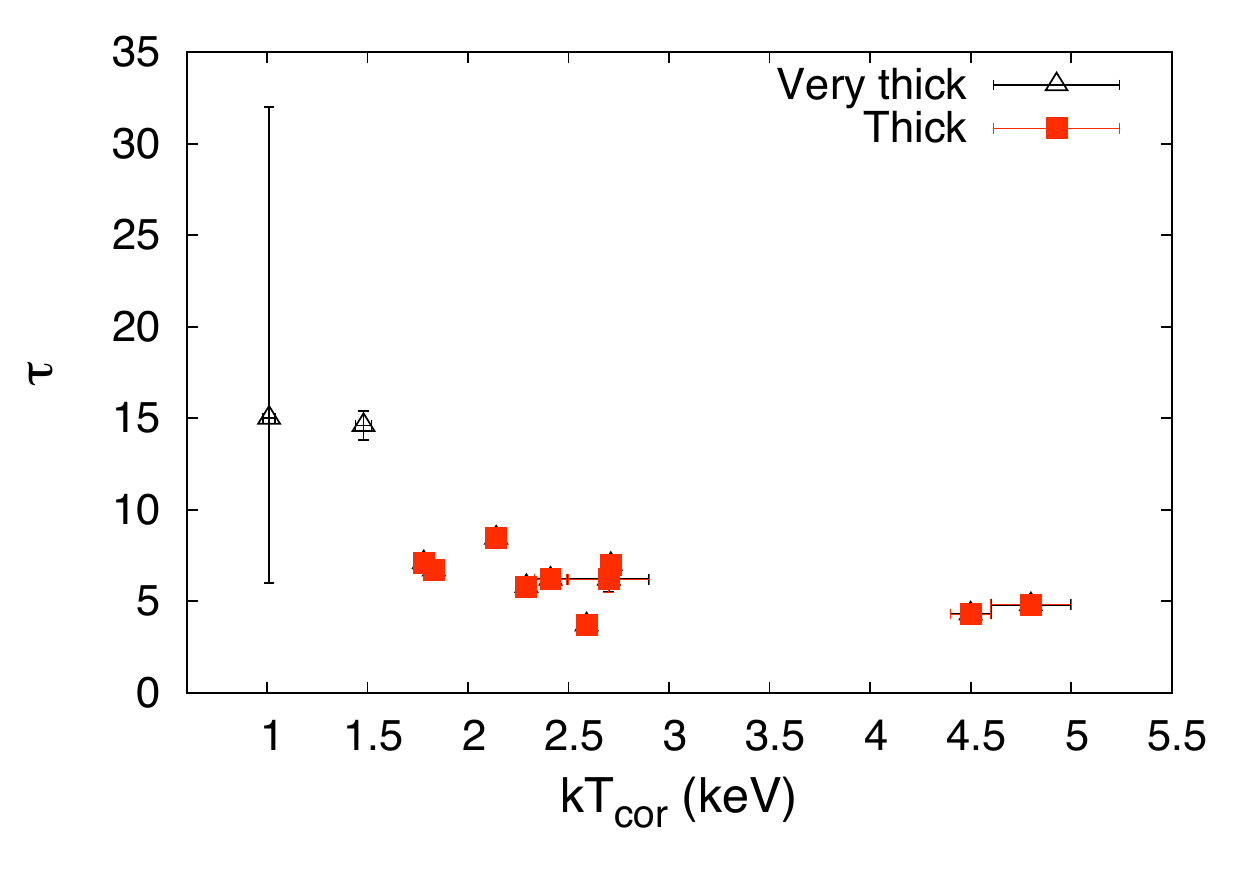}}
 \subfigure{\includegraphics[height=6.5cm,width=8.3cm]{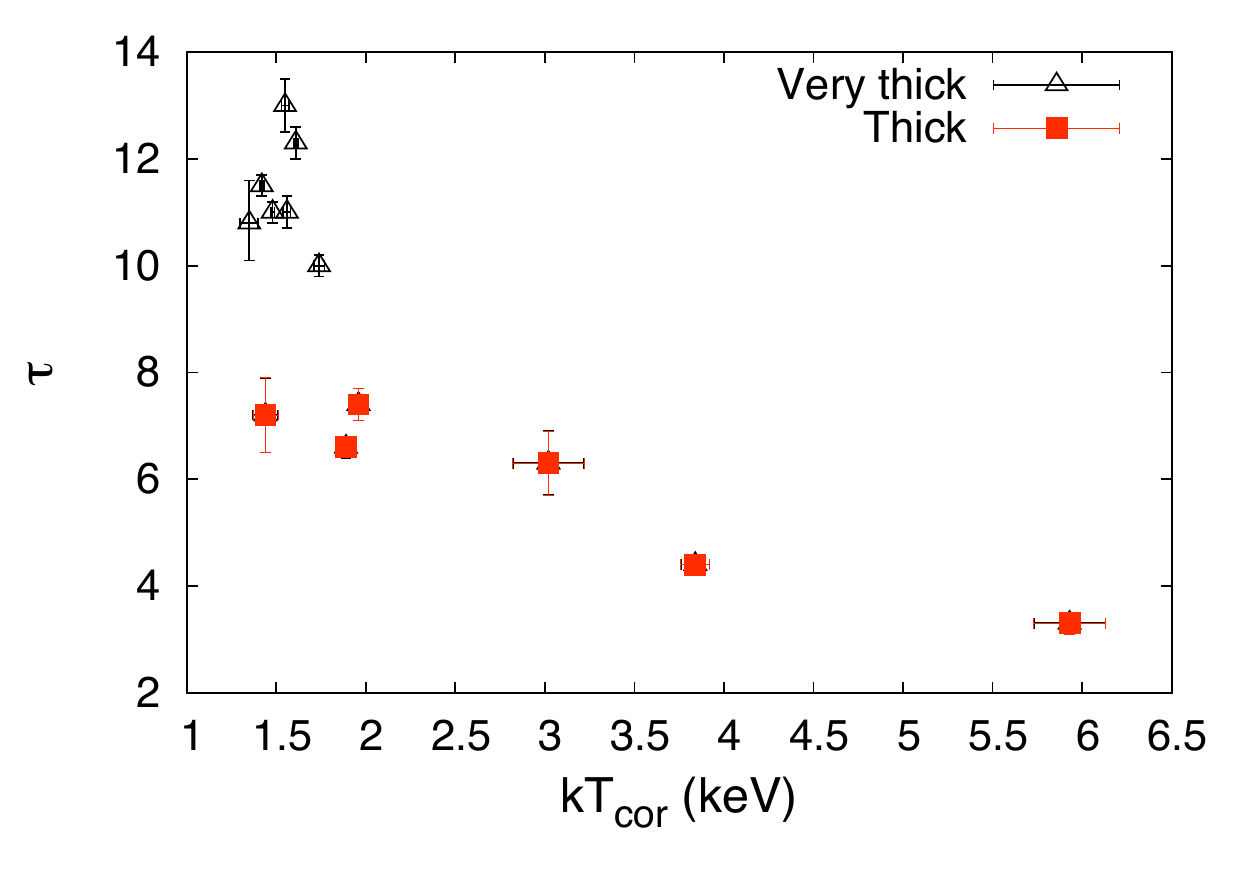}}
\caption{Optical depth $\tau$ versus temperature of the corona $kT_{cor}$ for the X-ray
spectral fits of X-1 ({\it left}) and X-2 ({\it right}) with a \textit{diskbb+comptt} model.  Observations that have a very thick corona are represented with {\it black} triangles while observations with a thick corona are represented as {\it red} squares. 
}
\label{figts}
\end{figure*}

\section{Data reduction}
\label{sect2}

We re-analyzed all the available {\it XMM-Newton} spectra of the two ULXs hosted in the spiral galaxy NGC 1313,
X-1 and X-2, with homogeneous criteria. The 17 observations span a time interval of six years, from 17 October
2000 to 16 October 2006, but three observations were excluded because of high flares contamination. 
Also the observation of October 2000 was excluded from the analysis because the calibration before 
December 2000 may be incomplete. 
Data were reduced using SAS v. 9.0.0. EPIC-MOS and EPIC-pn spectra were extracted selecting the good time intervals
with a background count rate not higher than 0.45 count s$^{-1}$ in the energy range $10-12$ keV. We set
`FLAG=0' in order to exclude the events at the CCD edge and the bad pixels. Spectra were extracted from events with
$PATTERN \leq 4$ for EPIC-pn (which allows for single and double pixel events) and $PATTERN \leq 12$ for EPIC-MOS (which allows
for single, double, triple and quadruple pixel events). We used $35"$ and $30"$ circular extraction regions 
for X-1 and X-2, respectively. For the background we chose a $65"$ extraction circular region on the same CCD chip where 
the source is located. If the sources were on or near a CCD gap, no spectra were extracted. 
For X-2 the off-axis angle is quite similar for most of the observations (not much larger than $0.4'$ for all but
three observations). For X-1, 10 observations have off-axis angles that differ no more than $1'$, while
the remaining three show a more significant variation. However, their spectra do not seem to present any peculiarity
possibly associated to variations in the fraction of encircled energy.
RGS spectra were extracted using the \textit{rgsproc} task with the option \textit{spectrumbinning=lambda}.  In
this way it turns out to be possible to combine the spectra of different observations. The EPIC spectra were
grouped with a minimum of 25 counts per channel, while RGS spectra with 50 counts per channel.

All the spectral fits were performed using XSPEC v. 12.5.1. To improve the counting statistics, whenever possible,
we fitted the EPIC-pn and EPIC-MOS spectra simultaneously. EPIC spectral fits were performed in the 0.3-10.0 keV
energy range, while RGS fits were limited to the 0.45-2.0 keV energy range. For each instrument, a multiplicative
constant was introduced to account for possible residual differences in the instrument calibrations. The constant 
of the EPIC-pn data set was fixed equal to 1, while the other two are allowed to vary. In general, the differences 
among the three instruments are not higher than 10$\%$.

\begin{figure}
 \subfigure{\includegraphics[height=6.3cm,width=8.3cm]{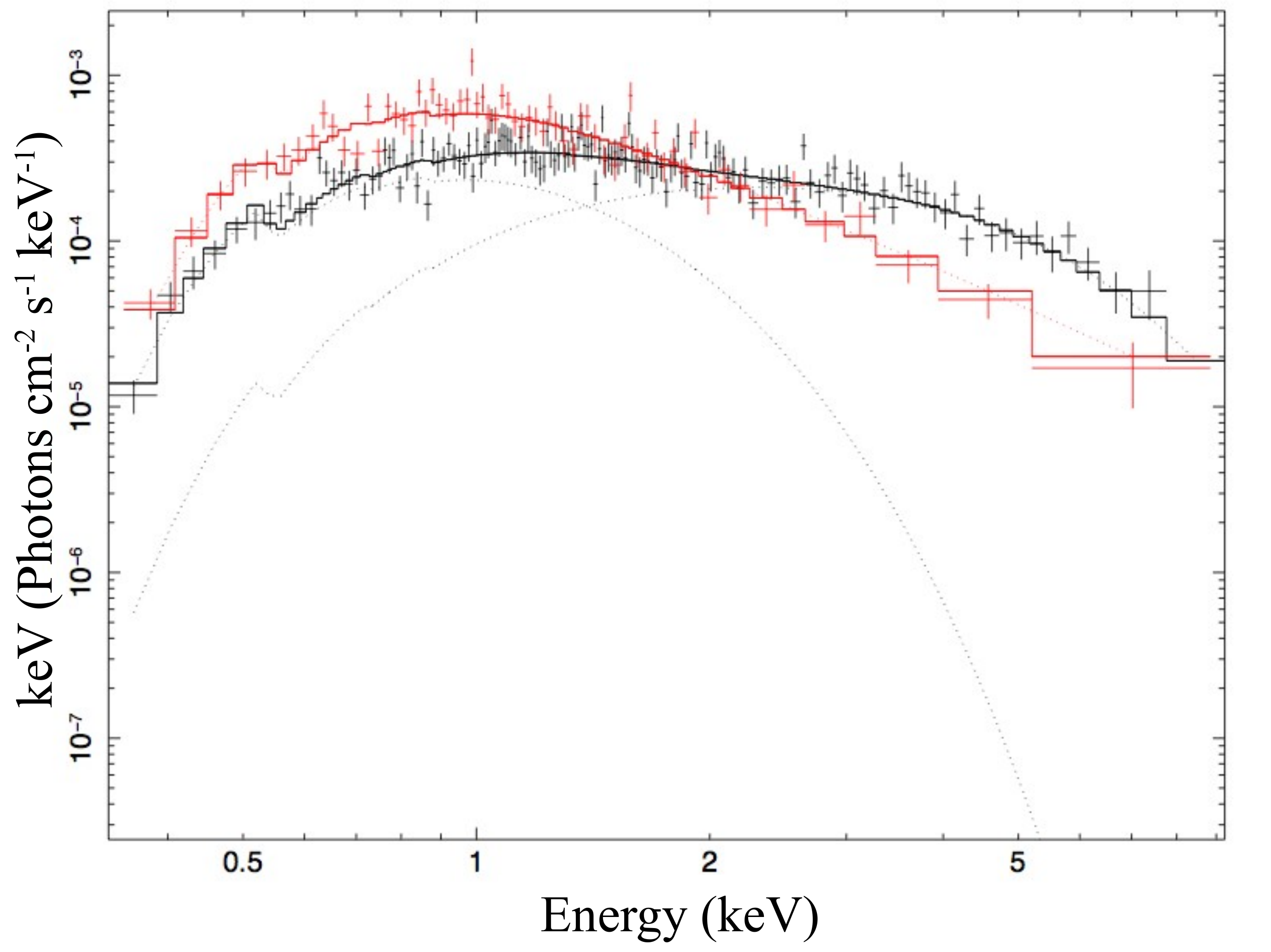}}
 \subfigure{\includegraphics[height=6.5cm,width=8.3cm]{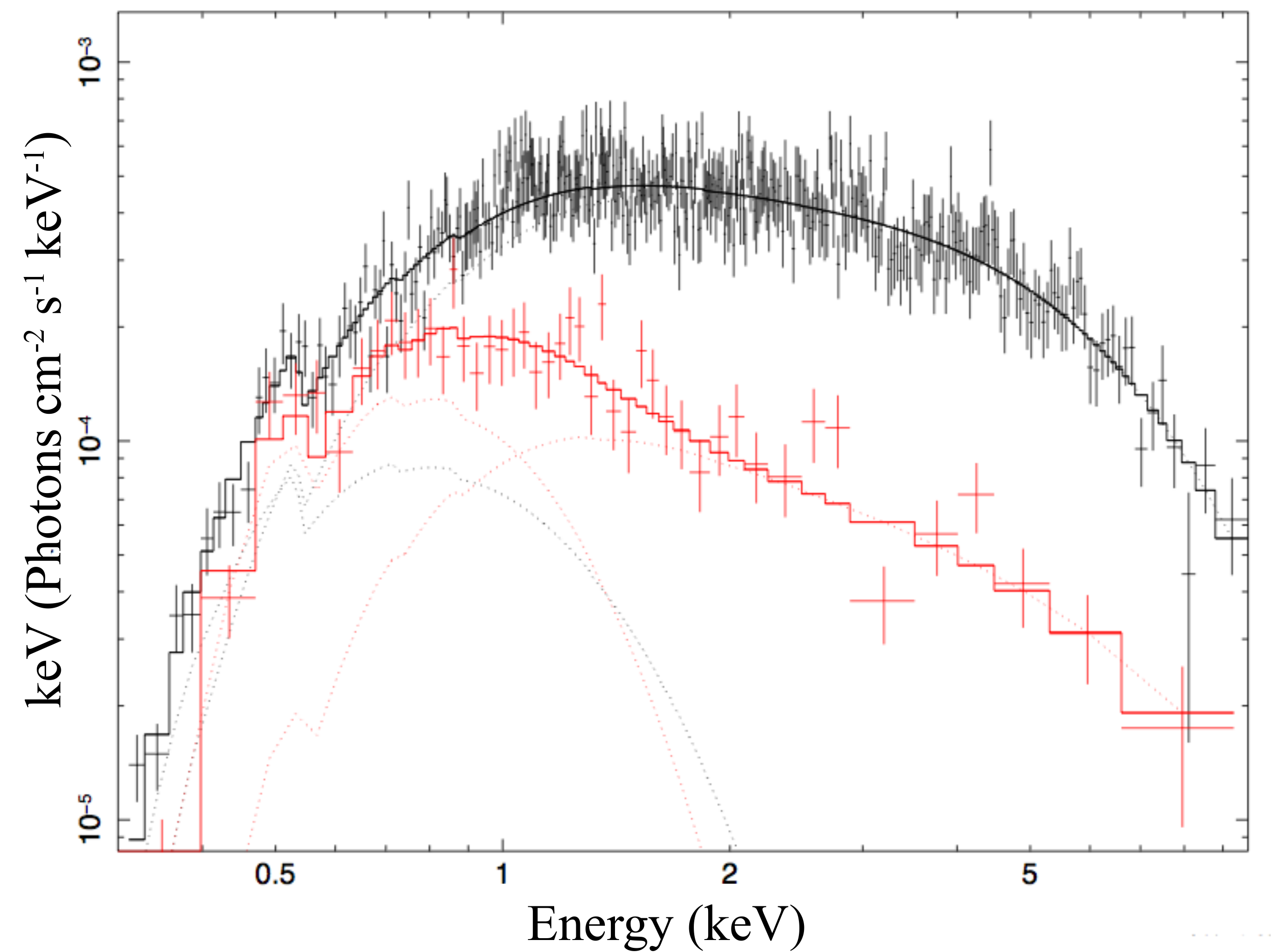}}
\caption{X-ray spectra (energy$\times$photons) of X-1 (top) and X-2 (bottom) for observations that have a very thick corona ({\it black}) and a thick corona ({\it red}). In both cases the comparison is between observations 
$\#9$ and $\#12$, respectively.}
\label{figsp}
\end{figure}

\begin{figure}
 \includegraphics[height=6.6cm,width=8.4cm]{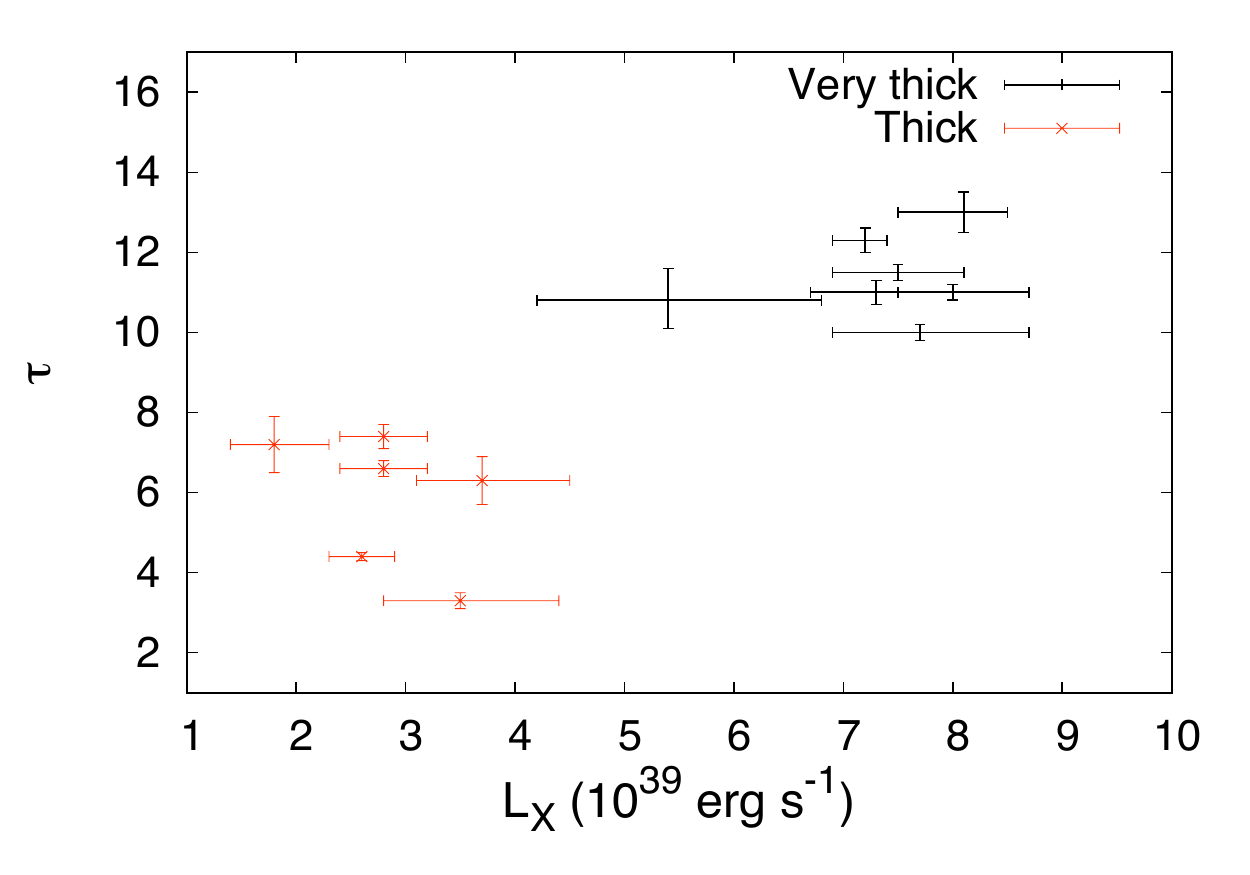}
\caption{Optical depth $\tau$ versus unabsorbed total luminosity (in the 0.3-10 keV range) for X-2. 
The {\it black} points represent the very thick corona state while the {\it red} crosses
are the thick corona state.}
\label{figtd4}
\end{figure}

\begin{table*}
\footnotesize
\begin{center}
   \caption{Best fitting spectral parameters of NGC1313 X-1 and X-2 in different observations obtained with the absorbed \textit{diskbb+comptt} model.}
   \label{ccd1}
   \begin{tabular}{l c c c c c c c c}
\hline
   & &  &  &  & NGC1313 X-1 &  &  &  \\ 
\hline
No. & Date & $N_H$$^a$ & $kT_{disc}$$^{b,c}$ & $kT_{cor}$$^d$ & $\tau$$^e$ & $L_{X}$ [0.3-10 keV] $^f$ & $L_{disc}$ [0.3-10 keV]$^g$ & $\chi^2/dof$ \\ 
&& ($10^{21}$ cm$^2$) & (keV) & (keV) & & ($10^{39}$ erg s$^{-1}$) & ($10^{39}$ erg s$^{-1}$) & \\
\hline
1&11/25/2003 & $0.2_{-0.2}^{+0.2}$	& $0.80_{-0.01}^{+0.01}$ & $9.03_{-5}^{+400}$ & $16.2_{-2}^{+2}$ & $10.0_{-0.3}^{+0.3}$ & $5.0_{0.4}^{1.2}$. & 21.33/38\\
2&12/21/2003 & $1.9_{-0.1}^{+0.1}$ & $ 0.199_{-0.005}^{+0.005}$ & $1.83_{-0.02}^{+0.02}$ & $6.7_{-0.1}^{+0.1}$ & $11.5_{-0.9}^{+0.9}$ & $0_{0}^{+3e36}$ & 195.47/226\\ 
3&12/23/2003 & $4.0_{-0.2}^{+0.2}$	& $0.229_{-0.004}^{+0.004}$ & $2.41_{-0.08}^{+0.08}$ & $6.2_{-0.3}^{+0.3}$ & $14.2_{-2.2}^{+2.5}$ & $4.6_{0.9}^{0.8}$.  & 64.38/75\\
4&12/25/2003 & $4.3_{-0.4}^{+0.4}$     &  $0.179_{-0.006}^{+0.005}$ & $2.7_{-0.2}^{+0.2}$ & $6.2_{-0.7}^{+0.7}$ & $14.9_{-4.8}^{+5.9}$ & $6.6_{2}^{2.4}$ & 12.88/12\\  
5&01/08/2004 & $2.7_{-0.1}^{+0.1}$	&  $0.244_{-0.003}^{+0.003}$ & $4.5_{-0.1}^{+0.1}$ & $4.3_{-0.2}^{+0.2}$ & $10.8_{-1.1}^{+1.3}$ & $2.7_{-0.3}^{+0.4}$ &  184.56/176\\ 
6&01/16/2004 & $1.3_{-0.1}^{+0.1}$ & $ 0.226_{-0.006}^{+0.006}$ & $2.28_{-0.04}^{+0.04}$ & $5.8_{-0.1}^{+0.1}$ & $9.3_{-0.6}^{+1.8}$ & $0_{0}^{1.6e38}$ & 194.71/197\\
7&05/01/2004 & $3.2_{-0.2}^{+0.2}$	&  $0.216_{-0.003}^{+0.003}$ & $4.8_{-0.2}^{+0.2}$ & $4.8_{-0.3}^{+0.3}$ & $8.0_{-1.3}^{+1.8}$ & $2.4_{0.4}^{0.6}$ & 100.97/91\\
8&06/05/2004 & $1.87_{-0.07}^{+0.07}$	&  $0.242_{-0.003}^{+0.003}$ & $1.78_{-0.02}^{+0.02}$ & $7.1_{-0.1}^{+0.1}$ & $15.0_{-1}^{+1}$ & $1.3_{0.2}^{0.2}$ & 511.38/543\\
9&08/23/2004 & $2.3_{-0.1}^{+0.1}$ & $0.158_{-0.002}^{+0.002}$ & $2.59_{-0.03}^{+0.03}$ & $3.70_{-0.07}^{+0.07}$ & $5.4_{-0.5}^{+0.6}$ & $0_{0}^{1.4e38}$ & 233.91/232\\ 
10&11/23/2004 & $2.2_{-0.1}^{+0.1}$ &  $0.367_{-0.003}^{+0.003}$ & $1.48_{-0.04}^{+0.04}$ & $14.6_{-0.8}^{+0.8}$ & $8.2_{-0.9}^{+1}$ &	$3.7_{0.3}^{0.3}$ & 165.41/184\\
11&02/07/2005 & $2.91_{-0.08}^{+0.09}$ & $0.208_{-0.002}^{+0.002}$ & $2.71_{-0.05}^{+0.05}$ & $7.0_{-0.2}^{+0.2}$ & $8.9_{-0.9}^{+1}$ & $2.7_{-0.3}^{+0.3}$ & 313.10/305\\
12&03/06/2006 & $2.3_{-0.1}^{+0.1}$ &  $0.459_{-0.004}^{+0.004}$ & $1.01_{-0.03}^{+0.03}$ & $15_{-9}^{+17}$ & $4.6_{-0.5}^{+0.6}$ & $2.5_{-0.2}^{+0.1}$ & 168.29/139\\
13&10/16/2006 & $2.67_{-0.002}^{+0.002}$ & $0.222_{-0.001}^{+0.001}$ & $2.14_{-0.01}^{+0.01}$ & $8.44_{-0.06}^{+0.06}$ & $7.1_{-0.2}^{+0.2}$ & $2.1_{-0.1}^{+0.1}$ & 1490.95/1420\\
\hline 
\hline
& &  &  &  & NGC1313 X-2 &  &  &  \\ 
No. & Date & $N_H$$^a$ & $kT_{disc}$$^b$ & $kT_{cor}$$^c$ & $\tau$$^d$ & $L_{X}$ [0.3-10 keV] $^e$ & $L_{disc}$ [0.3-10 keV]$^f$ & $\chi^2/dof$ \\ 
&& ($10^{21}$ cm$^2$) & (keV) & (keV) & & ($10^{39}$ erg s$^{-1}$) & ($10^{38}$ erg s$^{-1}$) & \\
\hline
1&11/25/2003 & $0.6_{-0.2}^{+0.2}$	& $0.44_{-0.01}^{+0.01}$ & $1.35_{-0.05}^{+0.05}$ & $10.8_{-0.7}^{+0.8}$ & $5.4_{-1.2}^{+1.4}$ & $12_{-4}^{+3}$& 58.72/62\\
2&12/21/2003 & $1.6_{-0.1}^{+0.1}$ & $ 0.380_{-0.004}^{+0.004}$ & $1.56_{-0.02}^{+0.02}$ & $11.0_{-0.3}^{+0.3}$ & $7.3_{-0.6}^{+0.7}$ & $14_{-2}^{+1}$ &  386.05/395\\
3&12/23/2003 & $0.5_{-0.1}^{+0.1}$	& $0.27_{-0.01}^{+0.01}$ & $1.74_{-0.03}^{+0.03}$ & $10.0_{-0.2}^{+0.2}$ & $7.7_{-0.8}^{+1}$ & $0_{0}^{6.4e+37}$  & 245.35/248\\
4&12/25/2003 & $1.2_{-0.2}^{+0.2}$ & $0.469_{-0.006}^{+0.006}$ & $3.02_{-0.2}^{+0.2}$ & $6.3_{-0.6}^{+0.6}$ & $3.7_{-0.6}^{+0.8}$ &$14.4_{-8}^{+2}$ &  99.68/120\\ 
5&01/08/2004 & $1.4_{-0.1}^{+0.1}$  &  $0.266_{-0.005}^{+0.005}$ & $1.89_{-0.04}^{+0.04}$ & $6.6_{-0.2}^{+0.2}$ & $2.8_{-0.4}^{+0.4}$ & $4.5_{-0.9}^{+1}$ &  175.27/201\\ 
6&01/16/2004 & $2.1_{-0.2}^{+0.2}$ & $ 0.186_{-0.005}^{+0.005}$ & $5.93_{-0.2}^{+0.2}$ & $3.3_{-0.2}^{+0.2}$ & $3.5_{-0.7}^{+0.9}$ & $4.7_{-1.7}^{+2.3}$ & 84.94/85\\
7&05/01/2004 & $1.0_{-0.3}^{+0.3}$	 &  $0.38_{-0.01}^{+0.01}$ & $1.44_{-0.07}^{+0.07}$ & $7.2_{-0.7}^{+0.7}$ & $1.8_{-0.4}^{+0.5}$ & $6.3_{-1.4}^{+1.7}$ & 55.41/46\\
8&06/05/2004 & $1.1_{-0.1}^{+0.1}$ &  $0.238_{-0.006}^{+0.006}$ & $1.48_{-0.01}^{+0.01}$ & $11.0_{-0.2}^{+0.2}$ & $8.0_{-0.5}^{+0.7}$ & $0_{0}^{+4.8e39}$ & 514.96/507\\
9&08/23/2004 & $3.0_{-0.1}^{+0.1}$ & $0.208_{-0.002}^{+0.002}$ & $1.96_{-0.05}^{+0.05}$ & $7.4_{-0.3}^{+0.3}$ & $2.8_{-0.4}^{+0.4}$ & $12_{2}^{2}$ & 150.72/131\\ 
10&11/23/2004 & $2.3_{-0.09}^{+0.1}$ & $0.226_{-0.002}^{+0.002}$ & $3.84_{-0.08}^{+0.08}$ & $4.4_{-0.1}^{+0.1}$ & $2.6_{-0.3}^{0+.3}$ &	$7.0_{0.9}^{1.1}$ & 210.89/237\\
11&02/07/2005 & $1.69_{-0.08}^{+0.09}$ & $0.633_{-0.004}^{+0.004}$ & $1.55_{-0.02}^{+0.02}$ & $13.0_{-0.5}^{+0.5}$ & $8.1_{-0.6}^{+0.4}$ & $28_{-1}^{+2}$ &  490.92/486\\
12&03/06/2006 & $1.54_{-0.08}^{+0.08}$	&  $0.288_{-0.006}^{+0.006}$ & $1.42_{-0.01}^{+0.01}$ & $11.5_{-0.2}^{+0.2}$ & $7.5_{-0.6}^{+0.6}$ & $5.1_{-0.9}^{+1}$ &  577.03/603\\
13&10/16/2006 & $1.89_{-0.04}^{+0.04}$ & $0.61_{-0.02}^{+0.02}$ & $1.61_{-0.01}^{+0.01}$ & $12.3_{-0.3}^{+0.3}$ & $7.2_{-0.3}^{+0.2}$ & $28_{-0.2}^{+1.1}$ &  841.83/854\\
\hline
   \end{tabular} 
 \end{center}
$^a$ Column density; $^b$ Inner disc temperature; $^c$ The seed photons temperature $T_0$ is assumed to be equal to $T_{disc}$.
$^d$ Temperature of the corona; $^e$ Optical depth of the corona; $^f$ Unabsorbed total x-ray luminosity in the 0.3 -10 keV range; 
$^g$ Unabsorbed disc luminosity in the 0.3-10 keV range. \\
\end{table*}

\normalsize

\section{X-ray spectral fits}
\label{sect3}

\subsection{Comptonization plus multicolor blackbody disc}

We analyzed 13 out of the 17 {\it XMM-Newton} observations of X-1 and X-2 adopting an absorbed multicolor 
blackbody disc plus Comptonization component, modelled with \textit{diskbb}+\textit{comptt} 
\citep{mitsuda84,titarchuk94} in XSPEC. This model was successfully adopted in previous
investigations to describe phenomenologically the spectra of ULXs (\citealt{stobbart06,gladstone09,feng09}). 
Although several of our observations show acceptable fits using only the \textit{comptt} component, 
in some cases adding the \textit{diskbb} component leads to a significant improvement in the fit.
Therefore, in order to perform a comparison within the framework of a unique spectral model,
here we assume the \textit{diskbb}+\textit{comptt} as reference model for all the observations
and check its consistency on the basis of the variability patterns of its spectral parameters.
The interstellar absorption was modelled with the \textit{tbabs}
model in {\it XSPEC}. We fixed the Galaxy column density along the line of sight at $3.9\times 10^{20}$
cm$^{-2}$ \citep{dickey90} and added a free absorption component to model the local absorption near the
source. 

The results of the spectral fits with the \textit{diskbb}+\textit{comptt} model for the 13
{\it XMM-Newton} observations of X-1 and X-2 are reported in Table~\ref{ccd1}. 
We tied the temperature of the disc ($T_{disc}$) to that of the seed photons ($T_0$) for comptonization. We will comment 
on this choice at the end of this Section.
The reported errors are at the 90\% confidence level for one interesting parameter.
We noted that there are often several local minima with close values of the $\chi^2$, 
sometimes with evidence for both a strong/warm and a weak/cool (or no) disc.
After a careful inspection of the $\chi^2$ surface, we found the absolute minima 
for each observation reported in Table~\ref{ccd1}. However, one should be aware that
the actual uncertainty on the disc parameters caused by the topology of the $\chi^2$ surface is at least
a few times larger than the formal error reported in the Table. 
We tried to adopt also more physical models, such as the \textit{eqtherm} \citep{coppi01} and the \textit{DKBBFTH} \citep{done06}.
We found that the counting statistics of most of the spectra is inadequate to constrain the parameters of \textit{eqtherm} 
and \textit{DKBBFTH} (see also below).

\begin{figure*}
\begin{center}
\subfigure{\includegraphics[height=5.9cm,width=7.8cm]{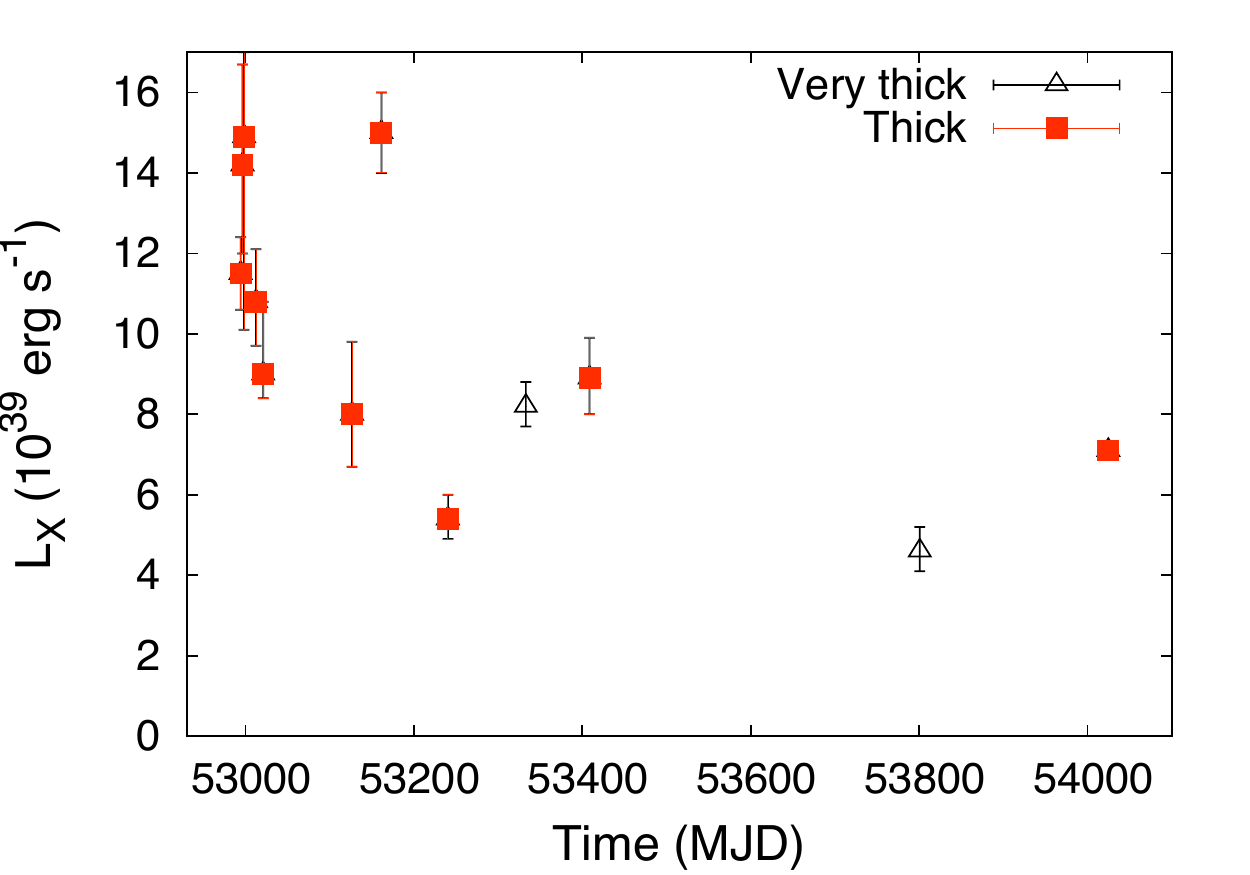}}
\subfigure{\includegraphics[height=5.9cm,width=7.8cm]{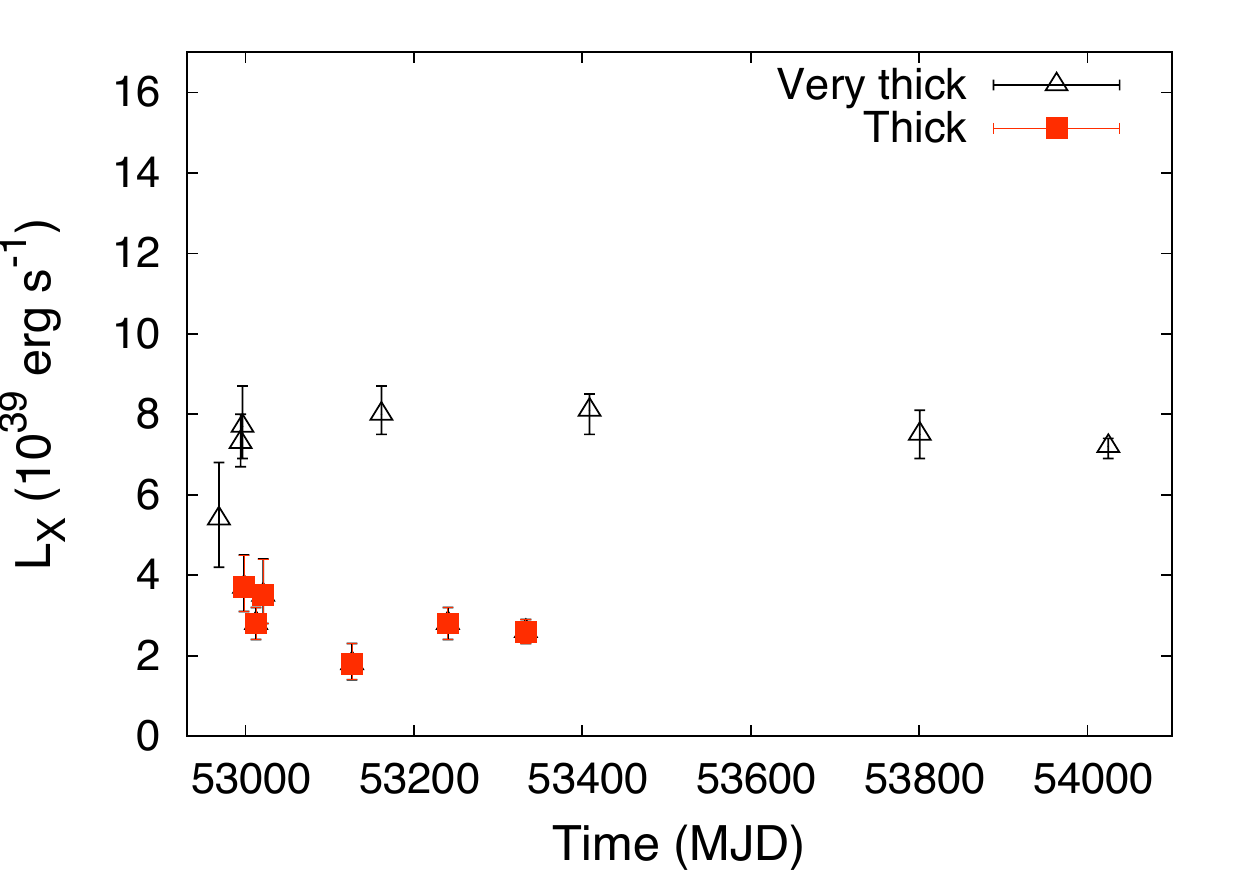}}
\caption{X-ray unabsorbed luminosity evaluated in the 0.3-10 keV energy band
(a distance of 3.7 Mpc was assumed; \citealt{tully88}) for all the {\it XMM-Newton} 
observations of X-1 ({\it up-left}) and X-2 ({\it up-right}). Observations with a very thick corona are 
represented with {\it black} triangles while observations with thick corona are represented as {\it red} 
squares. 
All the (unabsorbed) luminosities are evaluated in the 0.3-10 keV energy band. }
\label{figlc}
\end{center}
\end{figure*}

The longest observation of X-1 has a short segment of $\sim 3.5$ ks within the first 13 ks 
that shows some intrinsic variability (a slight systematic hardening at high energies) and was then
removed from the analysis.
We note also that observation \#1 has a very low value of $N_H$ compared to the other observations 
and a low counting statistics. It converges towards a local minimum with a very strong disc component
and with parameters unlike those of all the other spectra. 
For this reason, we excluded it from the following analysis.

Figure~\ref{figts} shows a plot of the optical depth $\tau$ versus the temperature of the corona $kT_{cor}$ 
obtained from the best fits of both sources.
There appears to be well defined locii in the $kT_{cor}$-$\tau$ plane, indicating the existence
of a somewhat ordered behaviour in the spectral variability of the corona. The spectra of X-2 populate
two distinct regions characterized by very large optical depths ($\tau \ga 10$) and low
temperatures ($kT_{cor} \sim$ 1.5 keV), on one side, and smaller optical depths ($\tau \la 8$)
and a range of temperatures ($kT_{cor} \sim$ 1.5-6 keV), on the other side.
In both cases the corona turns out to be optically thick. In the following we refer to these 
two regions as  ``very-thick"\ and ``thick"\ corona states, respectively. 
X-1 may also be interpreted within the same framework. An analysis of the 
X-ray flux variability shows that both sources tend to be significantly more variable when 
the coronae are thicker (fractional variability $\ga$9-10\%, see table \ref{ccd8}), which would strengthen the
analogies between their behaviours. However, the clustering of the 
observations in the $kT_{cor}$-$\tau$ plane of X-1 is not so well defined and the region with
high optical depths is populated by only two observations. Furthermore, although not included
in Figure~\ref{figts} because of the low counting statistics, our interpretation is not
consistent with the spectral paremeters inferred from observation \#1. 
So, present data are not sufficient to establish whether X-1
has indeed a bimodal behaviour similar to that shown by X-2 or the changes in the 
optical depth should be interpreted in a different way (e.g. as due to different
variability patterns or accretion geometries; \citealt{feng06,dewangan10}). 
In the following we will continue to distinguish between the observations of X-1 
with a very thick corona ($\tau \ga 10$) and a thick corona ($\tau \la 8$), being 
aware that they may represent physically different states with respect to those of X-2.

\begin{table*}
\begin{center}
   \caption{Fractional variability of NGC 1313 X-1 and X-2.}
   \label{ccd8}
   \begin{tabular}{l c c c c }
\hline 
& & \textit{Very Thick corona} & & \\
\hline
& X-2 (Obs. 13) & & X-1 (Obs. 12)$^c$ &  \\
Energy band$^a$ & counts $s^{-1}$ & $F_{var}$(per cent)$^b$& counts $s^{-1}$ & $F_{var}$(per cent)$^b$  \\
\hline
0.2-10 keV & $0.978\pm 0.005$ & $9.77\pm 0.07$ &  $1.47\pm 0.09$ & $14.6\pm 0.6$\\
\hline
\hline 
& & \textit{Thick corona} & & \\
\hline
& X-2 (Obs. 10) & & X-1 (Obs. 13)$^{c,e}$ &  \\
Energy band$^a$ & counts $s^{-1}$ & $F_{var}$(per cent)$^b$& counts $s^{-1}$ & $F_{var}$(per cent)$^b$  \\
\hline
0.2-10 keV& $0.377\pm 0.006$ & $<9.3^d$ & $0.77\pm 0.07$ & $2.3^{+0.9}_{-1.9}$ \\
\hline
\end{tabular}
\end{center}
$^a$ The adopted energy band is for a direct comparison with \citet{dewangan10}
$^b$ Calculated from the background subtracted EPIC-pn light curve, with 200 s time bins.
$^c$ From \citet{dewangan10}. 
$^d$ 3 $\sigma$ upper limit. 
$^e$ Fractional variability computed for the whole observation, including the short segment of 3.5 ks that we left out from the spectral analysis.
\end{table*}

An example of the spectral shapes when the two sources are in different positions on the 
$kT_{cor}$-$\tau$ plane is shown in Figure~\ref{figsp}. Again there are analogies and differences
between X-1 and X-2. In both cases
the spectra for very high coronal depths are bell-shaped, with a clear turn-over at $\ga 3-4$ keV (e.g.
\citealt{stobbart06}), whereas the spectra in the low-$\tau$ region are steeper and do not show strong evidence 
of curvature at high energies. Only for X-2 the two spectral states appear to correlate with total luminosity, 
the very-thick corona state being more luminous (see Figure~\ref{figtd4}). In X-1 there is no significant
dependence of the spectral shape on $L_X$. This is evident also from the behaviour of the spectra shown in Figure~\ref{figsp}.
While for X-2 the thick corona spectra stay always below those in the very thick corona state, for X-1 there
is a sort of crossing/pivoting point of the observed spectra (see also \citealt{kajava09}) at $\sim 2$ keV. 
Therefore, the total counts 
in the thick corona state of X-2 are clearly smaller than those in the very thick corona state, while in X-1 the deficit 
of photons observed at low energies in spectra with larger optical depths is compensated by the excess of photons 
at high energies.
Figure~\ref{figlc} shows the light curves of X-1 and X-2 computed from the best fitting
\textit{diskbb+comptt} model (a distance of 3.7 Mpc was assumed; \citealt{tully88}). On average, X-1 has a
luminosity ($\sim 10^{40}$ erg s$^{-1}$) higher than that of X-2 ($\sim 5 \times 10^{39}$ erg s$^{-1}$).  
Variability of a factor $\sim 3$ and $\sim 5$ is observed for X-1 and X-2, respectively.
We further investigated the behaviour of the soft component in both sources.
The discs are soft or warm, with temperatures of $\sim 0.2-0.5$ keV for X-1 and $\sim 0.2-0.6$ keV for X-2.
Three spectra of X-1 are consistent with zero normalization (or absence) of the soft component.
In eight observations the luminosity of the soft component is $\sim 10^{39}$
erg s$^{-1}$ and it represents a significant fraction of the total flux ($\ga 30$\%). 
Excluding the spectral fits that return zero normalization of the soft component,
there is no evidence of correlation or anti-correlation between the disc (or total)
luminosity and the inner disc temperature.
In X-2 the luminosity of the soft component is a significant fraction ($\ga 30$\%) of the total 
luminosity in six observations. Two observations have normalization consistent with zero.
Excluding them, the disc luminosity appears to show a weak power-law correlation with the inner
temperature, $L_{\mathrm{disc}}\propto T_{disc}^{1.2\pm 0.3}$.
However, the correlation is uncertain, as the two observations that substantiate it
(those with higher $kT_{disc}$) have also rather shallow
minima in $\chi^2$, admitting both a strong/warm and a weak/cool disc fit with 
close values of the $\chi^2$. No correlation is found using the total luminosity.

\subsection{Effects of varying the ratio of seed photons temperature to the disc temperature}

The comptonizing coronae of the \textit{comptt} model turn out to be optically 
thick. This poses a problem, as in these physical conditions the disc underneath 
the corona is masked by it. Therefore, the temperature of the disc component refers 
to the outer visible part of the disc, while the temperature of the seed photons 
$T_0$ is not necessarily equal to $T_{disc}$. 
For this reason, \citet{gladstone09} adopted the {\it DKBBFTH} model in which the 
corona is assumed to cover the inner disc. 
We tried to apply the same model to our spectral sequence, but found that it converges
to plausible physical values of the parameters only for the highest quality spectra. 
We then tried to repeat our analysis
with the {\it diskbb}+{\it comptt} model and disconnecting the two temperatures,
but this leads to difficulties in finding a global minimum (see also the analysis
of NGC 5204 X-1 in \citealt{feng09}). Finally, we decided to test our results
tying the two temperatures with a fixed proportionality constant.

If the corona is optically thick and is absorbing a constant fraction $f$ of the accretion power, 
the actual inner disc temperature $T_1^{'}$ is lower than the inner temperature of the disc
in absence of the corona $T_1$, as $T_1^{'}=T_1 (1-f)^{1/4}$ (e.g. \citealt{gladstone09}).
If $R_1$ and $R_T$ are the inner disc radius and the truncation radius of the corona respectively,
from the relation $T(R)\approx T_1(R/R_1)^{-3/4}$, valid for a disc in absence of the corona,
we obtain $T_1\approx T_T(R_T/R_1)^{3/4}$. Thus, we have $T_1^{'}\approx T_T(R_T/R_1)^{3/4} 
(1-f)^{1/4}$. Assuming that $f\la 90$\% and that the corona is compact ($R_T\la 4 R_1$), 
the inner disc temperature $T_1^{'}$, which is also the seed photons temperature, 
is larger than the temperature of the outer visible disc $T_T$, and $1 \la T_1^{'}/T_T=T_0/T_{disc} \la 2.5$.
We then attempted to perform some additional fits with two fixed values (1.5, 2) of the 
ratio $\delta = T_0/T_{disc}$ as representative of this situation.

We found that the spectral fits with a {\it diskbb}+{\it comptt} model
and $\delta=1.5, 2$ are statistically acceptable and the coronae are still optically 
thick, with the inferred optical depth weakly depending on $T_0/T_{disc}$.
The parameters of the corona change by no more than $\sim 60$\%, with typical
variations of $\la 20-30$\%. However, the dependence of spectral states on the total luminosity
that characterizes the behaviour of X-2 for equal temperatures is lost.
For X-1, the temperature of the disc continues to not correlate 
with the luminosity, as found for equal temperatures.
For X-2, varying $\delta$ the very-thick state becomes less populated, as some observations
previuosly in that state move to the thick corona state. 
For $\delta=1.5$ there is a weak correlation between disc 
luminosity and temperature ($L_{disc}\propto T_{in}^{1.4\pm 0.4}$),
while if $\delta=2$ the correlation disappears. At variance with the correlation found for equal
temperatures, the correlation for $\delta=1.5$ is not critically dependent only on two observations.

These results do not change significantly for slightly more extended coronae.
On the other hand, if the corona is very extended and optically thick, the value of $\delta$ 
may become so large that the disc component falls essentially outside the {\it XMM-Newton} bandpass
(unless the disc-corona coupling is very strong). In these conditions some spectra are no 
longer well fitted by the model.

\begin{figure}
 \includegraphics[height=6.5cm,width=8.4cm]{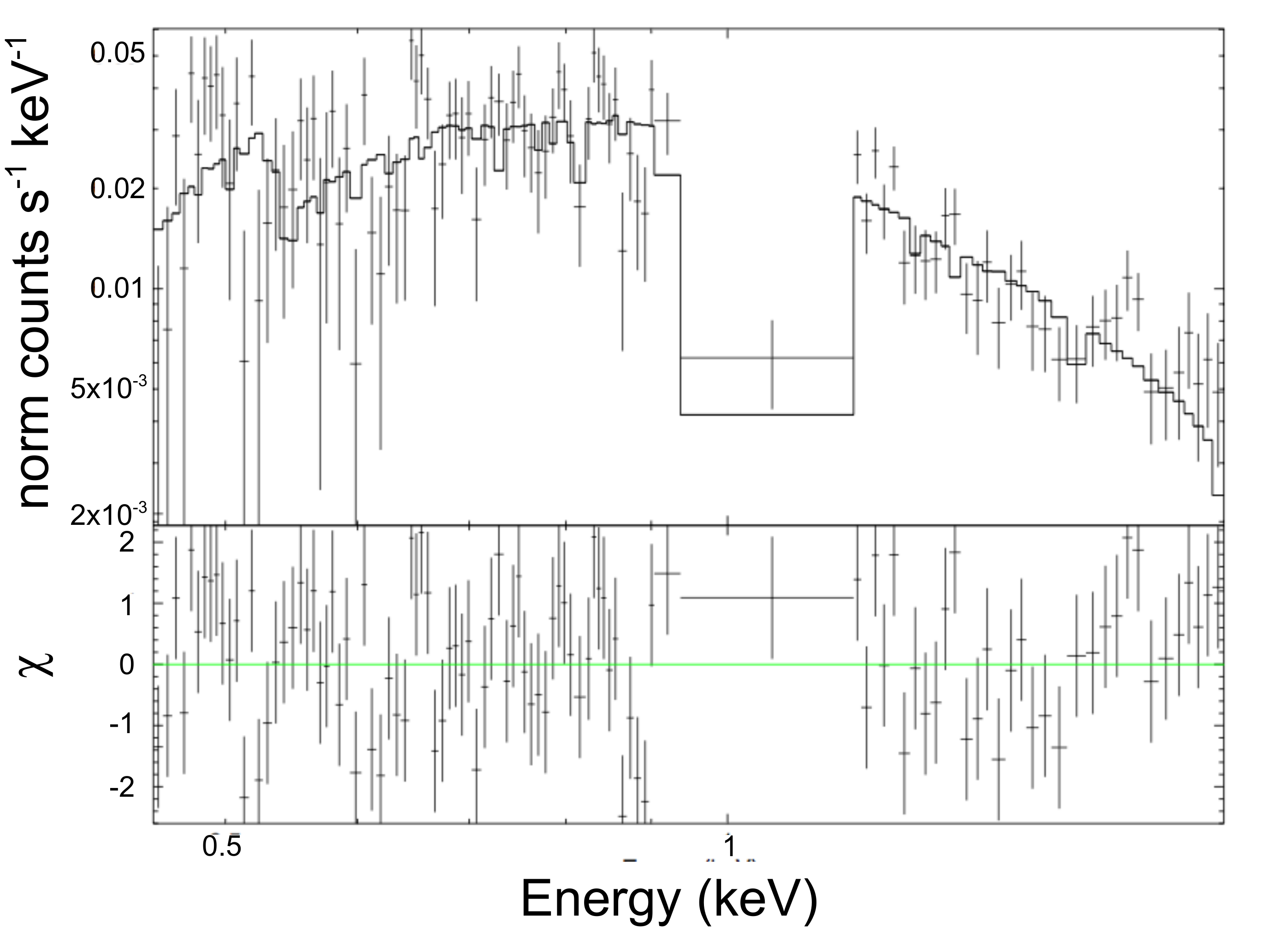}
\caption{Stacked RGS1 spectrum of X-1, along with its best fitting model (see text for details).}
\label{figtd2}
\end{figure}

\begin{table}
  \begin{center}
   \caption{Abundances inferred from the Oxygen K-shell photoionization edge (0.538 keV)
   from the X-ray spectral fits of NGC 1313 X-1 and X-2.}
   \label{ccd5}
   \begin{tabular}{l c c}
\hline 
& NGC 1313 X-1 & \\
\hline
& $\tau^a$ & $12+\log(O/H)^b$  \\
\hline
RGS (observation 13) & $0.75_{-0.25}^{+0.29}$ & $8.7_{-0.2}^{+0.1}$ \\
EPIC (observation 13) & $0.75_{-0.03}^{0.03}$ & $8.72_{-0.02}^{+0.02}$ \\
RGS (stacked) & $0.53_{-0.23}^{+0.26}$ & $8.7_{-0.2}^{+0.3}$ \\
\hline
\hline
& NGC 1313 X-2 & \\
\hline
& $\tau^a$ & $12+\log(O/H)$ \\
\hline
EPIC (observation 13) & $0.46_{-0.04}^{+0.05}$ & $8.64_{-0.05}^{+0.06}$ \\
\hline
\end{tabular}
\end{center}
$^a$ Absorption depth at the threshold energy. \\
$^b$ For the solar abundance we assume $12+\log(O/H)=8.92$ ($Z_\odot=0.02$).
\end{table}

\section{Chemical abundance estimates}
\label{sect4}

We analyzed the EPIC-pn and RGS data of all the {\it XMM-Newton} observations of X-1 and X-2 
in an attempt to use them for determining the chemical abundances in the local source environment.
Following \citet{winter07}, for the EPIC-pn 5000 and 40000 counts are necessary 
to observe the Oxygen K-shell and Iron L-shell absorption edges at 0.538 keV and 0.851 keV, respectively. 
Observation $\#13$ has $\sim$ 58000 (53000) EPIC-pn counts for X-1 (X-2), and hence we can perform a
chemical abundance analysis on the EPIC-pn spectrum similar to that presented in \citet{winter07}.
In the spectral fits
the \textit{tbabs} absorption model is replaced with \textit{tbvarabs} that allows to vary the chemical abundances 
(and grain composition). We set alternatively the abundance of Oxygen or Iron to zero. The
spectrum was then fitted with the EPIC continuum model in Table~\ref{ccd1} (keeping all parameters, but 
normalizations, fixed) plus an absorption edge, that accounts for the observed absorption feature. 
The parameters of the edge are then used to compute the abundance. 

Using this approach, we found that
the O and Fe abundances inferred from the EPIC-pn spectrum of X-1 are consistent with
a sub-solar metallicity environment ($Z\sim 0.6 Z_\odot$ from the O edge; Table~\ref{ccd5}). 
The $\chi^2$ of the fit with the absorption edge is 1488 for 1423 dof, while that without the edge is
1759 for 1424 dof. For X-2 we find a subsolar abundance for Oxygen ($Z \sim 0.5 Z_\odot$; see again Table~\ref{ccd5})
and an Iron abundance consistent with zero. The fits returns $\chi^2=839$ (857 dof) with the O edge and 
$\chi^2=953$ (858 dof) without it.

We tried to analyze also the RGS data of X-1 and X-2 using the same technique. 
As the sources are faint, the RGS net count rate is quite low
($\sim 1.5\times 10^{-2}$ count s$^{-1}$). Therefore, only the last observation (observation $\#13$, 122 ks)
reached a reasonable counting statistics in the RGS for the brightest ULX (X-1), while no useful
analysis could be performed on X-2. We tentatively identified two features in absorption 
in the RGS spectrum of X-1, associated to O I (0.535 keV) and Fe I (0.709 keV). There
may be also two other lines in emission at the characteristic energy of O VIII K$_{\alpha}$ 
(0.653 keV) and Si K$_{\alpha}$ (1.748 keV), but their significance is very low.
From the Oxygen edge we found an abundance below solar ($Z\sim 0.6 Z_\odot$) for the absorbing 
material towards X-1 (see Table~\ref{ccd5}), although the statistical improvement obtained 
including the absorption edge is small.

We tried to improve the analysis stacking together the RGS spectra in such a way
to increase the counting statistics. We used only the RGS1 spectra, because our
best diagnostic is the Oxygen K-shell photoionization edge at 0.538 keV and
the response of the RGS2 has some problems precisely in this energy range (0.5-0.6 keV).
The stacked spectra of X-2 have not enough counting statistics for
a meaningful analysis. For X-1, we combined spectra together using the 
\textit{rgscombine} command, that appropriately accounts for the response matrices and 
backgrounds of the different observations.
We had to exclude several observations ($\#2,3,4,5,6,7,11,12$) that caused technical problems in the processing with XSPEC. 
Observation $\#1$ was also excluded for the reason explained in the previous Section.
The final combined RGS spectrum of X-1 has a total exposure time of 168 ks and $\sim 2861$ net counts. 
We fitted it with a \textit{diskbb+comptt} model with parameters fixed and equal to the mean values obtained 
from the EPIC spectral fits of X-1, setting the Oxygen abundance to zero and adding an absorption edge
(Figure~\ref{figtd2}). We found that the Oxygen abundance is again subsolar (Table~\ref{ccd5}). 
The $\chi^2$ of the fit with the absorption edge is 139 for 103 dof, while that without the edge is
146 for 104 dof. 
However, the actual abundance is rather sensitive to the temperature of the soft component and its uncertainty becomes large
if $kT_{disc}$ is included in the spectral fit.

\section{Discussion}
\label{sect5}

The analysis of all the {\it XMM-Newton} observations shows that the spectra of X-1 and X-2 can be well
reproduced by a Comptonization model plus a soft disc component in which the coronae are always
optically thick. Both sources appear to show a well defined 
behaviour in the optical depth ($\tau$) versus corona temperature ($kT_{cor}$) plane. For X-2 we 
clearly identified two
states that characterize the spectral variability and appear to have also different short term 
variability properties: a ``very-thick" (more variable) corona state in which
$\tau \ga 10$ and $kT_{cor} \sim$ 1.5 keV, and a ``thick-corona'' state in which $\tau \la 8$
and $kT_{cor} \sim 1.5-6$ keV. We note that a morphological classification in terms
of a harder brighter state and a softer dimmer state, based on a single power-law fit, 
has already been proposed for X-2 \citep{feng06}. Here we offer a physical explanation of 
these two states in terms of varying parameters of an optically thick corona.
The behaviour of X-1 may be interpreted within the same framework but, with presently
available data, the observed changes in the coronal optical depth could be explained
also in a different way (e.g. as due to different variability patterns or accretion 
geometries; \citealt{feng06,dewangan10}).

For X-2, which is on average less luminous than X-1, the two spectral states of the corona appear 
to correlate with luminosity. The optical depth of the corona increases as the total luminosity goes up,
as expected if the corona responds rapidly to an increment in the instantaneous accretion rate.
The behaviour of the disc component is also different in the two sources. While in X-1 the luminosity
and temperature of this component do not correlate, for X-2 we find that $L_{disc}\propto T_{disc}^{1.2\pm 0.3}$.
However, as already mentioned, this result is uncertain as the two spectra that substantiate the
correlation admit both a strong/warm and a weak/cool disc fit with close values of the $\chi^2$.

A spectral analysis based on the assumption that the seed photon temperature ($T_0$) is equal to the disc
temperature ($T_{disc}$) has some inconsistencies because, if the corona is optically thick, the innermost part 
of the accretion disc is actually not visible and, hence, it is not necessarily true that $T_0=T_{disc}$.
For this reason, we repeated our analysis with the \textit{diskbb}+\textit{comptt} model 
tying the two temperatures with a proportionality constant, that was fixed assuming a compact corona
energetically coupled to the disc (disconnecting the two temperatures
leads to difficulties in finding a global minimum because of the low counting statistics of several
observations). We find changes in the corona parameters of $\la 60$\%, with typical
variations of $\la 20-30$\%. However, the dependence of spectral states on the total luminosity
that characterizes the behaviour of X-2 for equal temperatures is lost.
While for X-1 the temperature of the disc continues not to correlate 
with the luminosity, for X-2 the correlation persists up to values of $T_0/T_{disc} \la 1.5$, consistent
with strong disc-corona coupling ($f\sim 90$\%).

We emphasize that the model adopted in the present investigation is not entirely
physically consistent. Besides the issues of the relation between $T_0$ and $T_{disc}$
that we tried to address as described above, there may be other caveats, such as
the input photons not being a mono-temperature Wien distribution (as assumed in \textit{comptt}),
or the disc structure being different from a standard one at high accretion rates,
or the origin and precise location of the thick corona/wind being unkown.
However, our purpose was not to adopt the most physically consistent spectral 
model for ULXs, which does not exists yet, but a simplified one that can 
reflect the underlying physics and can be tested on the basis of the observed
spectral variability patterns, limited by the current data quality.

A full explanation of the observed spectral variability patterns of X-1 and X-2 is beyond the scope
of the present investigation and appears not easy. Here we simply propose some interpretations
within the framework of recent work on the subject (e.g. \citealt{feng09,gladstone09,dewangan10,vierdayanti10}).
X-1 shows higher average isotropic luminosity ($\sim 10^{40}$ erg s$^{-1}$) and smaller variability. 
The different spectral shapes do not correlate with luminosity, and the temperature and luminosity 
of the soft component do not vary together.
These findings are consistent with X-1 being in the ultraluminous regime \citep{gladstone09}, accreting at 
super-Eddington rates and launching powerful winds from a disc embedded in an optically thick corona. 
The accretion disc may be there, but in a different physical regime and covered 
by a wind/corona, while the soft component may actually represent emission from the wind itself. 
The larger X-ray fractional variability observed when the wind/corona is optically thicker may
be consistent with increasing obscuration of the source caused by the turbolent wind itself
at higher accretion rates (\citealt{middleton11}a,b).

As far as X-2 is concerned, it behaves in a way similar to X-1 on the $\tau$-$kT_{cor}$ plane and also the fractional X-ray variability of the two sources appears to be comparable.
Hence the corona may be in a similar physical state. Again, this may
be consistent with the picture in which we are seeing the two sources under a high 
inclination angle and the temporal variability may be produced by blobs in the wind 
that intersect our line of sight to the hottest central regions of the source 
(\citealt{middleton11}a,b). However, as noted above, X-2
has a lower average luminosity and, for $T_0\la 1.5 T_{disc}$, it shows a weak 
correlation between the luminosity and inner temperature of the disc component.
This may be interpreted as X-2 being in a less extreme regime, in which the average 
accretion rate is at around or slightly above the Eddington limit. The soft spectral
component may then physically represent a true accretion disc which is partly
visible and has a characteristic temperature in the range 0.2-0.6 keV. Clearly, 
some time-dependent interaction between the disc and the corona is likely to occur 
(e. g. a slight expansion on a dynamical timescale of the corona as the accretion rate
increases) and may be responsible for the observed slope of the temperature-luminosity 
relation ($L_{disc} \propto T_{disc}^{1.2}$), which is different from that of a standard 
disc with fixed inner radius ($L_{disc} \propto T_{disc}^4$). 
Clearly, as the corona is always optically thick, it is not possible to use the disc 
parameters for estimating the BH mass.

We note that the disc temperature-luminosity correlation in X-2 disappears using a simple 
power-law to describe the spectrum of the corona, as found by \citet{feng06} (and 
by \citet{feng09} for IC 342 X-1), probably because the power-law does not take into 
account the spectral curvature at high energies.
The correlation of the optical depth to the corona with luminosity that we found
for X-2 (assuming $T_0=T_{disc}$) is in agreement with the results obtained for Ho IX X-1 by \citet{vierdayanti10}. 
Also IC 342 X-1 shows some hint of an increase in the coronal depth as the X-ray
luminosity increases (\citealt{feng09}). These similarities in spectral variability patterns
appear to be consistent with the proximity of these three ULXs in the spectral sequence proposed 
by \citet{gladstone09}.

We used the RGS high spectral resolution to attempt an estimate of the metallicities of the local environments
of X-1 and X-2. Because of the low signal-to-noise ratio, only the last, longest observation of X-1 could be
used. The analysis was performed also on the EPIC spectrum. The metallicity in the X-1 environment is
consistent with being below solar.
The last EPIC spectrum of X-2 was also analyzed with the same method and suggests subsolar metallicity. 
In an attempt to increase the significance of our results we also stacked together the RGS spectra of some
observations of X-1 and performed measurements of the metallicity on the stacked spectrum.
Also in this case the Oxygen abundance turns out to be below solar, although it is rather sensitive to 
the temperature of the soft component. 
Our estimates are in agreement with the abundance measurements from HII regions in NGC 1313 that give 
values that are all subsolar (\citealt{pilyugin01, hadfield07,ripamonti10}), but are smaller than
the slightly supersolar metallicity of the X-1 enviroment found by \citet{winter07}. The difference
with the latter Authors may be due to the fact that we analyzed an observation with higher counting 
statistics and adopted a different spectral model.

A couple of ULXs show periodic intensity variations in X-rays which are considered as signatures of the orbital
period (\citealt{k1,k2,k3,strohmayer09}). X-ray spectra seem to change regularly
with orbital phase. It is possible that these phase related variations may affect any result obtained from
snapshot observations such as those presented here. Furthermore, as mentioned above, the observed spectral 
variability patterns of X-1 and X-2 are rather complex and their interpretation does not appear directly 
comparable to that of Galactic XRBs.
For these reasons, the acquisition of new, high quality spectra through a dedicated X-ray monitoring 
programme is definitely needed. 

\section*{Acknowledgements} We would like to thank Richard Mushotzky for suggesting the stacking of RGS 
spectra, Chris Done, Tim Roberts and Jeanette Gladstone for helpful remarks on the adopted spectral models,
and Emanuele Ripamonti for his suggestions on the metallicity estimates.
We thank also the two referees of this paper for useful comments. We acknowledge financial support through INAF grant PRIN-2007-26 and ASI/INAF grant n. I/009/10/0.

\bsp
\label{lastpage}


\begin{thebibliography}{1}
\bibitem[\protect\citeauthoryear{Begelman}{2006}]{begelman06} Begelman M.C., 2006, ApJ, 643, 1065
\bibitem[\protect\citeauthoryear{Belczynski et al.}{2010}]{belczynski10} Belczynski K., Bulik T., Fryer C. L., Ruiter A., Valsecchi F., Vink J. S., Hurley J. R., 2010, ApJ, 714, 1217
\bibitem[\protect\citeauthoryear{Colbert \& Mushotzky}{1999}]{colbert99} Colbert E. J. M., \& Mushotzky R. F., 1999, ApJ, 519, 89
\bibitem[\protect\citeauthoryear{Coppi}{2001}]{coppi01} Coppi P.S., 2001, MNRAS
\bibitem[\protect\citeauthoryear{Dewangan et al.}{2010}]{dewangan10} Dewangan G. C., Misra R., Rao A. R., Griffiths R. E., 2010, MNRAS, 407, 291
\bibitem[\protect\citeauthoryear{Done \& Kubota}{2006}]{done06}Done C., Kubota A., 2006, MNRAS, 371, 1216
\bibitem[\protect\citeauthoryear{Dickey et al.}{1990}]{dickey90}Dickey J.M., Lockman F.J., 1990, ARA\&A, 28, 215
\bibitem[\protect\citeauthoryear{Fabbiano}{1989}]{fabbiano89}Fabbiano G.,1989, ARA\&A, 27,87
\bibitem[\protect\citeauthoryear{Feng \& Kaaret}{2005}]{feng05} Feng H., Kaaret P., 2005, ApJ, 633, 1052
\bibitem[\protect\citeauthoryear{Feng \& Kaaret}{2006}]{feng06} Feng H., Kaaret P., 2006, ApJ, 650, 75
\bibitem[\protect\citeauthoryear{Feng \& Kaaret}{2009}]{feng09} Feng H., Kaaret P., 2009, ApJ, 696, 1712
\bibitem[\protect\citeauthoryear{Fryer}{1999}]{fryer99} Fryer C. L., 1999, ApJ, 522, 413
\bibitem[\protect\citeauthoryear{Gierlinski et al.}{1999}]{gier99} Gierlinski M., Zdziarski A. A., Poutanen J., Coppi P. S., Ebisawa K., Johnson W. N., 1999, MNRAS, 309, 496
\bibitem[\protect\citeauthoryear{Gierlinski \& Done}{2004}]{gier04} Gierlinski M., Done C., 2004, MNRAS, 347, 885
\bibitem[\protect\citeauthoryear{Gladstone et al.}{2009}]{gladstone09} Gladstone J.C., Roberts T.P., Done C., 2009, ApJ, 397, 1836
\bibitem[\protect\citeauthoryear{Goncalves \& Soria}{2006}]{goncalves06} Goncalves A.C., Soria R., 2006, MNRAS, 371, 673
\bibitem[\protect\citeauthoryear{Hadfield et al.}{2007}]{hadfield07} Hadfield L. J., Crowther P.A., 2007, MNRAS, 381, 418
\bibitem[\protect\citeauthoryear{Kaaret et al.}{2006a}]{k1} Kaaret P., Simet M. G., Lang C. C., 2006a, Science 311, 491
\bibitem[\protect\citeauthoryear{Kaaret et al.}{2006b}]{k2} Kaaret P., Simet M. G., Lang C. C., 2006b, ApJ 646, 174
\bibitem[\protect\citeauthoryear{Kaaret \& Feng}{2007}]{k3} Kaaret P., Feng H., 2007, ApJ 669, 106
\bibitem[\protect\citeauthoryear{Kajava \& Poutanen}{2009}]{kajava09} Kajava J., Poutanen J., 2009, MNRAS, 398, 1450
\bibitem[\protect\citeauthoryear{King et al.}{2001}]{king01} King A.R., Davies M. B., Ward M. J., Fabbiano G., Elvis M., 2001, ApJ, 552, L109
\bibitem[\protect\citeauthoryear{King}{2009}]{king09} King A. R., 2009, MNRAS, 393, L41
\bibitem[\protect\citeauthoryear{Linden et al.}{2010}]{linden10} Linden T., et al. 2010, ApJ, 725, 1984
\bibitem[\protect\citeauthoryear{Mapelli et al.}{2009}]{mapelli09} Mapelli M., Colpi M., Zampieri L. 2009, MNRAS, 395, L71
\bibitem[\protect\citeauthoryear{Mapelli et al.}{2010}]{mapelli10} Mapelli M., Ripamonti E., Zampieri L., Colpi M., Bressan A. 2010, MNRAS, 408, 234
\bibitem[\protect\citeauthoryear{Middleton et al.}{2011}]{middleton11} Middleton M.J., Sutton A. D., Roberts, T. P., 2011, MNRAS, tmp,1209
\bibitem[\protect\citeauthoryear{Middleton et al.}{2011}] {middleton11} Middleton M.J., Roberts T.P., Done C., Jackson F.E., 2011, MNRAS, 411, 644
\bibitem[\protect\citeauthoryear{Miller et al.}{2002}]{miller02}Miller M. C., Hamilton D. P., 2002, MNRAS, 330, 232
\bibitem[\protect\citeauthoryear{Miller et al.}{2003}]{miller03} Miller J. M., Fabbiano G., Miller M. C., Fabian A. C., 2003, ApJ, 585, 37
\bibitem[\protect\citeauthoryear{Miller et al.}{2004}]{miller04} Miller J. M., Fabian A. C., Miller M. C., 2004, ApJ, 607, 931
\bibitem[\protect\citeauthoryear{Mizuno et al.}{2007}]{mizuno07} Mizuno T., et al., PASJ, 59, 257
\bibitem[\protect\citeauthoryear{Mitsuda et al.}{1984}]{mitsuda84} Mitsuda K., Inoue H., Koyama K., Makishima K., Matsuoka M., Ogawara Y., Suzuki K., Tanaka Y., Shibazaki N., Hirano T., 1984, PASJ, 36, 741 
\bibitem[\protect\citeauthoryear{Mucciarelli et al.}{2007}]{mucciarelli07} Mucciarelli P.,Zampieri L., Treves A., Turolla R., Falomo R., 2007, ApJ, 658, 999
\bibitem[\protect\citeauthoryear{Pilyugin}{2001}]{pilyugin01} Pilyugin L.S., 2001, A\&A, 369, 594
\bibitem[\protect\citeauthoryear{Pintore \& Zampieri}{2011}]{pintore11} Pintore F., Zampieri L., 2011, AN, 332, 337
\bibitem[\protect\citeauthoryear{Poutanen et al.}{2007}]{poutanen07} Poutanen J., et al., 2007, MNRAS, 377, 1187
\bibitem[\protect\citeauthoryear{Ripamonti et al.}{2010}]{ripamonti10} Ripamonti E., Mapelli M., Zampieri L., Colpi M.
\bibitem[\protect\citeauthoryear{Socrates \& Davis}{2006}]{socrates06} Socrates A., \& Davis S. W., 2006, ApJ, 651, 1049
\bibitem[\protect\citeauthoryear{Stobbart et al.}{2006}]{stobbart06} Stobbart A. M., Roberts T. P.,Wilms J., 2006, MNRAS, 368, 397
\bibitem[\protect\citeauthoryear{Strohmayer}{2009}]{strohmayer09} Strohmayer T. E.: 2009, ApJ 706, L210
\bibitem[\protect\citeauthoryear{Swartz et al.}{2008}]{swartz08} Swartz D. A., Soria R., \& Tennant A. F., 2008, ApJ, 684, 282
\bibitem[\protect\citeauthoryear{Titarchuk}{1994}]{titarchuk94} Titarchuk L., 1994, ApJ, 434, 570
\bibitem[\protect\citeauthoryear{Tully}{1988}]{tully88} Tully R. B., 1988, Nearby Galaxies Catalog. Cambridge Univ. Press, Cambridge
\bibitem[\protect\citeauthoryear{Vierdayanti et al.}{2010}]{vierdayanti10} Vierdayanti K., Done, C., Roberts T. P., Mineshige S., 2010, MNRAS, 403, 1206
\bibitem[\protect\citeauthoryear{Winter et al.}{2007}]{winter07} Winter L. M., Mushotzky R, F., Reynolds C. S., 2007, Apj, 655, 163
\bibitem[\protect\citeauthoryear{Zampieri \& Roberts}{2009}]{zampieri09} Zampieri L., Roberts T.P., 2009, MNRAS, 400, 677 
\end{thebibliography}
\end{document}